\documentstyle[prb,aps,twocolumn]{revtex}

\begin{document}
\flushbottom
\draft
\twocolumn[\hsize\textwidth\columnwidth\hsize\csname @twocolumnfalse\endcsname

\title{Low-energy quantum dynamics of atoms at defects.
Interstitial oxygen in silicon}
\author{Rafael Ram\'{\i}rez,$^1$ Carlos P. Herrero,$^1$ 
        Emilio Artacho,$^2$ and F\'elix Yndur\'ain$^2$}
\address{
   $^1$Instituto de Ciencia de Materiales, Consejo Superior de
   Investigaciones Cient\'{\i}ficas (C.S.I.C.),
   Campus de Cantoblanco, 28049 Madrid, Spain
\\
   $^2$Instituto de Ciencia de Materiales Nicol\'as Cabrera and
   Departamento de F\'{\i}sica de la Materia Condensada, C-III,
   Universidad Aut\'onoma de Madrid, 28049 Madrid, Spain}
\date{15 July 1996}
\maketitle
\begin{abstract}
The problem of the low-energy highly-anharmonic quantum dynamics of
isolated impurities in solids is addressed by using path-integral
Monte Carlo simulations. Interstitial oxygen in silicon is studied
as a prototypical example showing such a behavior.
The assignment of a "geometry" to the defect is discussed.
Depending on the potential (or on the impurity mass), there is a
"classical" regime, where the maximum probability-density for
the oxygen nucleus is at the potential minimum. There is another regime,
associated to highly anharmonic potentials, where this is not
the case. Both regimes are separated by a sharp
transition. Also, the decoupling of the many-nuclei problem into a
one-body Hamiltonian to describe the low-energy dynamics is studied.
The adiabatic potential obtained from the relaxation of all the other
degrees of freedom at each value of the coordinate associated to the
low-energy motion, gives the best
approximation to the full many-nuclei problem.
\end{abstract}
%
% PACS
%
% 61.72.Ji Defects in crystals: Point defects (interstitials ...)
% 63.20.Pw Lattice dynamics, localized modes
% 66.30.Jt Diffusion, migration, and displacement of impurities
%
%\pacs{PACS numbers: 63.20.Pw, 61.72.Ji, 66.30.Jt}
%
\pacs{ }

]

\section{INTRODUCTION}

   The dynamics of impurities or, more generally, atoms at point defects in
crystalline solids quite often gives rise to localized low-energy excitations,
typically in the far infrared (FIR) spectral region, displaying patterns that
reflect, in one way or another, a substantial deviation from the situation of
atomic nuclei harmonically vibrating around their potential 
minimum.\cite{bosomworth,gienger,muro} For example, interstitial H 
or Li around other substitutional impurities in Si and Ge are 
delocalized among several positions equivalent by symmetry, giving rise to FIR
excitations.\cite{complexes} Similar FIR excitation patterns
are found in glasses.\cite{glasses} Mixed crystals have recently
been shown to exhibit a similar behavior.\cite{sievers}

The substantial deviation from harmonicity in these systems, 
considering the quantum character of their dynamics, is of great
importance in their characterization.
On one hand, the quantum character prevents the 
assignment of a definite geometry to the defect
structure, the nuclei showing high probability for being found away from
the potential minimum location. On the other hand, the anharmonicity 
makes that the many-nuclei problem
cannot be easily decoupled as usually in terms of normal modes of vibration,
leaving open the question of what is the best decoupling approximation, that
accounts for the experimental observations of those low-energy excitations.

   Interstitial oxygen ($O_i$) in silicon also displays nontrivial 
quantum behavior.\cite{bosomworth,kaneta,ego} The oxygen atom is known 
to break a Si--Si bond, establishing two Si--O bonds in the form of an 
oxygen bridge. The motion of the oxygen atom and its neighbors in the 
direction of the axis of the original Si--Si bond can be described in terms of
harmonic vibrations, which account for some of the main features in the
infrared absorption spectrum of the center.\cite{ego} It is the motion
of the oxygen atom in the plane perpendicular to that axis what gives
rise to the non-trivial behavior, characterized by a peculiar FIR
spectrum.\cite{bosomworth} The symmetry of the center corresponds to the
$D_{3d}$ point group. This symmetry facilitates the a priori
partial decoupling
of the relevant degrees of freedom. It is this fact, together
with the availability of experimental and theoretical information,
what makes this system particularly suitable for analysis of
the questions raised above: (i) can we define a geometry for the center, and
(ii) what is the best way of decoupling 
the low-energy dynamics from the rest.

   The relevance of the geometry question has already been pointed out in the
literature.\cite{ego} It was observed that, for the effective potential
obtained from experiment,\cite{kaneta} the oxygen atom has
maximum probability-density at the bond-center (BC) site in spite of 
being the potential a
local maximum at that point. The arguments used there, however, are for
the effective one-particle Hamiltonian, and have to be checked for the
fully interacting problem.

   The importance of the decoupling problem is clear by the analysis of the 
experimental FIR spectra in terms of finding  an
effective one-particle potential in two dimensions, such that an
oxygen atom moving under that potential would reproduce the
vibrational frequencies observed in the
spectra. It is an axially-symmetric potential well with a local maximum at the
bond-center and a minimum around it, at 0.22 \AA\ off the BC site
(mexican-hat shape).\cite{kaneta} The energy barrier amounts to
$\approx 1$ meV. If we consider the fact that the oxygen atom is strongly
interacting with its silicon neighbors, the question arises: what is the
meaning of that potential? The multidimensional potential associated to all
the nuclear degrees of freedom of the system is well defined, but not a
two-dimensional one, unless a decoupling prescription is available.
In similar situations, when the harmonic decoupling is meaningless,
propositions for decoupling are found in the
literature ranging between two extremes. Arguing on the basis of the lightness
of the impurity atom, some

\twocolumn[\hsize\textwidth\columnwidth\hsize\csname @twocolumnfalse\endcsname

\parindent 10pt
\parskip 2pt

\noindent
authors\cite{denteneer} propose that the 
effective one-particle
potential for the motion of that impurity is the one obtained when the
host (heavier) atoms are kept fixed at some defined positions
(fixed-lattice potential; FLP,
hereafter): the effective slower motion of the host
atoms is presumed from mass considerations. The other extreme can be also 
stated in terms of the different times scales of the relative motions, but 
taking now into consideration the energies associated to those motions.
 In this latter case
it is the impurity (or some of its degrees of freedom) the slow moving, in
spite of its lightness. This argument leads to an adiabatic potential (AP)
that is obtained by allowing the host atoms to relax for each value of the
coordinates associated to the low-energy degrees of freedom.

   These questions are addressed in this paper by solving the full quantum
many-nuclei problem of interstitial oxygen in silicon
by means of path-integral (PI) Monte Carlo (MC) simulations.
This paper is divided up in the following manner. In Sec.\,II we
briefly describe the computational method and summarize the details
concerning the potential used. Sec.\,III is devoted to the results and
discussion. The geometry of the center $O_i$ is assessed
with the probability-density function of
the oxygen nucleus. The decoupling of the low-energy motion
is studied by comparing the results of the complete MC simulations with those
derived according to different decoupling criteria.
The paper closes with a summary (Sec.\,IV).

\section{COMPUTATIONAL METHOD}

\subsection{Path-integral simulations}

The PI MC method has become a
standard tool to study finite-temperature
properties of quantum systems. In this Section we 
briefly present the details relevant for the
presentation of our results. More complete descriptions of
the PI formalism can be found 
elsewhere.\cite{ceperley95,gillan90,takahashi84,freeman84,pollock84}
The canonical partition function of $P$
silicon nuclei plus the O impurity can be expressed with the
Trotter formula and the high-temperature
approximation for the density
matrix\cite{gillan90} as:

\begin{equation}
Z \approx \left( \frac{m_{\text O}}{m_{\text{Si}}} \right)^{3N/2}
 \left( \frac {Nm_{\text{Si}}}{2\pi\beta{\hbar}^2}\right) ^ {3(P+1)N/2}
 \int \prod_{j=1}^{N} \, d{\bf R}_j \;
 \exp \left\{ -\beta v_{\text {eff} }
                ({\bf R}_1,...,{\bf R}_N) \right\} \, .
\label{z}
\end{equation}
The index $j$ represents the path coordinate along a cyclic path, 
which has been decomposed into a number $N$
of discrete intervals (Trotter number).
${\bf R}_j$ is a vector in a $3(P+1)$-dimensional space,
whose components are the
Cartesian coordinates of the $P+1$ nuclei
$({\bf r}_{1,j};...;{\bf r}_{P+1,j})$.
The cyclic condition for the path
coordinate of each nucleus is expressed
as ${\bf R}_{N+1} = {\bf R}_{1}$.
The masses of the  host and impurity atoms are
$m_{\text{Si}}$ and $m_{\text O}$ , respectively; $\beta
= (k_{B} T)^{-1} $, and $k_{B}$ is the Boltzmann constant.
Eq. (\ref{z}) coincides with the canonical
partition function of a classical system
with the effective interaction potential:

\begin{equation}
v_{\text {eff}}({\bf R}_1,...,{\bf R}_N) =
\sum_{j=1}^{N}
               \left \{ A({\bf R}_j,{\bf R}_{j+1})
                      + N^{-1} V({\bf R}_j) \right \} ,
\label{veff1}
\end{equation}
where
\begin{equation}
A({\bf R}_j,{\bf R}_{j+1}) =
      \frac {N}{2 \beta^2 {\hbar }^2}
      \left \{ m_{\text O} ({\bf r}_{P+1,j+1} - {\bf r}_{P+1,j})^2 +
      \sum_{p=1}^{P} m_{\text {Si}} ({\bf r}_{p,j+1} - {\bf r}_{pj})^2
      \right \} \; .
\label{veff2}
\end{equation}
The index $p$ indicates the particle, and goes from 1 to $P$ for the
silicon atoms, and takes the value $P + 1$ for the impurity.
The function
$v_{\text {eff}}({\bf R}_1,...,{\bf R}_N)$ is the
interaction potential of a classical system composed of $P+1$ cyclic
chains {[}one per nucleus; a total of $N(P+1)$
classical particles{]} characterized
by a harmonic intrachain coupling with a force constant
$\kappa = {mN}/{\beta^2 {\hbar }^2}$ {[}first term on the
right-hand side of Eq. (\ref{veff1}); $m = m_{\text {Si}}$ or 
$m_{\text O}$, depending on the nucleus{]}.
Interchain coupling {[}second term on
the right-hand side of Eq. (\ref{veff1}){]}
is restricted to those
particles with the same index $j$, and
this interaction is equal to that
corresponding to the quantum particles,
$V({\bf R}_j)$, but renormalized by a factor $N^{-1}$ 
(inverse of the number of discrete
points along the path coordinate).
The approximated expression for $Z$ {[}Eq. (\ref{z}){]} becomes
exact in the limit $N \rightarrow \infty$, and is valid for
distinguishable particles.
This assumption of distinguishable particles
is reasonable for the statistics of Si nuclei,
since exchange effects are negligible.

Equilibrium properties of the quantum system    
can be derived 
by Metropolis Monte Carlo sampling of the multidimensional integral
associated to the
partition function given by Eq.
(\ref{z}).\cite{metropolis53,valleau91,binder88} A
simulation run proceeds via successive MC steps (MCS).
In each MCS, the nuclei coordinates ${\bf r}_{p,j}$ are updated according
to two different kinds of sampling

]

\clearpage
 
\noindent
schemes. The first one
consists on trial moves of the individual coordinates ${\bf r}_{p,j}$.
The trials are performed sequentially for every path coordinate $j$ and
every nucleus $p$. The second type of sampling ]
corresponds to
trial moves of the center of gravity of the cyclic paths, that are
carried out sequentially for every nucleus $p$
in the simulation cell. The number of MCS's
employed for
system equilibration was of $5 \times 10^3$,
while the calculation of ensemble average properties was performed
over $2 \times 10^5$ MCS's.
For the Si atoms we have employed the average
isotope mass of this element ($m = 28.086$ amu).
The number $N$ of discretized points for the
path coordinate was made
temperature dependent, and was taken as the integral number
closest to $2000/T$, a condition that guarantees convergence in the
total energy within a relative error smaller than 1\%. \cite{ramirez93}

\subsection{Potential}

The Monte Carlo simulations have been performed on a
$2\times2\times2$ supercell of the Si face-centered cubic (fcc) cell
containing 64 Si atoms and an oxygen impurity.
The simulation cell was subject to
periodic boundary conditions. The interaction between silicon atoms has
been described by the three-body potential developed by Stillinger
and Weber.\cite{stillinger85}
The potential between oxygen and silicon atoms
has been designed to reproduce qualitatively the main features
of the $O_i$ defect: (i) the overall geometry, O 
breaking a bond between two Si atoms; (ii) the observed low-energy (FIR) 
excitations\cite{bosomworth} (even though
the PI does not provide excited states, it does give internal energy
versus temperature, which, at low temperatures is mainly controlled by 
these low-energy excitations); (iii) the vibrations of the center at
higher frequencies, known by infrared absorption;\cite{ir} and (iv) the
main features of the potential obtained from first-principles
calculations, including bond lengths, bond angle, and Si relaxations.\cite{ego}
This potential for the interaction between O and Si atoms 
is a function of the coordinates of both
the impurity
(${\bf r}_{\text O})$, and the Si atoms (${\bf r}_1$ and ${\bf r}_2$)
coordinated to the oxygen:
\begin{equation}
V({\bf r}_1, {\bf r}_2, {\bf r}_{\text O}) =
     V_r(r_{1{\text {O}}}) + V_r(r_{2{\text {O}}}) + V_s(\alpha)
                 + V_l(\alpha) \; ,
\label{pot1}
\end{equation}
where ${\bf r}_{i{\text {O}}} = {\bf r}_{i} - {\bf r}_{{\text {O}}}$
($i$ =1, 2), and $\alpha$
is the angle Si--O--Si. The potential functions are:
\begin{equation}
V_r(r) = \frac{1}{2} k (r-r_e)^2   \; ,
\label{pot2}
\end{equation}
\begin{equation}
V_s(\alpha) = s_1 \sin^2 \alpha  + s_2  \sin^4 \alpha  \; ,
\label{pot3}
\end{equation}
\begin{equation}
V_l(\alpha) = l_1 ( \cos \alpha  - \cos {\alpha}_e )^2 \; ,
\label{pot4}
\end{equation}
with the following values of the constants:
$k=35.6$ eV {\AA}$^{-2}$,
$r_e = 1.629$ \AA, $s_1 = 1.49$ eV,
$s_2 = -0.7484$ eV,
$l_1 = 5.4$ eV, and ${\alpha}_e = 168^{\text {o}}$.
The local geometry of the defect complex
is shown in Fig.\,\ref{sketch}. The absolute potential minimum
corresponds to a geometry with
the oxygen nucleus located at the off-center position M,
at a distance of about 0.29 \AA\ from the Si--Si axis. The corresponding
Si--O distance is 1.52 \AA\ and the Si--O--Si angle is 158$^{\text
{o}}$. The
relaxation of the nearest Si atoms along the [111] crystal direction
amounts to 0.32 \AA. For comparison, the Si--O distance obtained from
total-energy Hartree-Fock calculations\cite{ego} of the $O_i$ defect
is 1.56 \AA, with an outwards
relaxation of the nearest Si atoms of 0.36 \AA\ each.
The vibrational frequency of the A$_{1g}$ mode at the $O_i$
center derived from our model potential is 587 cm$^{-1}$, while the
frequency reported in a cluster-Bethe-lattice investigation\cite{ego}
of the $O_i$ defect
was 569 cm$^{-1}$. This mode corresponds to atom
displacements along the Si--Si axis and has no infrared activity
because of symmetry.
The potential energy for the impurity located at the 
BC site is higher than that found for the absolute minimum.
This value agrees with that of $\sim$ 1 meV
corresponding to the model potential of
Yamada-Kaneta {\em et al.},\cite{kaneta} which was designed to
reproduce the spectrum of low-energy excitations of the $O_i$ center.
When the oxygen is located at the BC site, our parametrized potential
gives a relaxation of the nearest Si atoms of 0.35 \AA.

In Fig. \ref{potential}(a), a calculated potential energy surface  
is presented as a function of both the distance between the Si atoms 
nearest neighbors of O, $d$(Si--Si), and the distance from the oxygen 
atom to the bond-center site as oxygen moves 
in the plane perpendicular to the Si--Si axis,
$d$(O--BC). For every point in the figure, the positions of the
other Si atoms were relaxed. The curves obtained by sectioning the 
surface for fixed values of $d$(Si--Si) correspond to FLP's.
For distances $d$(Si--Si) larger than 3.04 \AA\ the
minimum energy is found when the oxygen atom is located at the BC site.
However, at smaller values of $d$(Si--Si), the minimum
is found for an off-center position.
In Fig. \ref{potential}(b) we present two sections of
the energy surface calculated
at representative distances $d$(Si--Si).
The broken line shows the FLP derived with the Si atoms
fixed at their relaxed positions for the absolute minimum of the 
potential, the oxygen impurity
being located at the off-center position M, with $d$(Si--Si) = 3 \AA\
(called FLP-1, hereafter). The energy barrier amounts to $\sim$ 24 meV. 
The solid line corresponds to a fixed-lattice potential (FLP-2) with the
hosts atoms fixed at their relaxed
positions for the O nucleus located at the BC site and
$d$(Si$_1$--Si$_2$) = 3.05 \AA. The
dotted line connecting the minima of both curves represents the adiabatic
potential (AP) characterized by an energy
barrier. 

\section{RESULTS AND DISCUSSION}

\subsection{Geometry}

   In order to define a geometry for the $O_i$ defect,
we have studied the
probability-density $\rho(r)$ of finding the oxygen nucleus at a
distance $r$ from the Si--Si axis along any direction
by PI MC simulations at 10 K. The full quantum treatment included
the O nucleus and the nearest and next-nearest Si neighbors
(i.e., a total of nine atoms),
while the remaining Si atoms in the simulation cell were
fixed at their relaxed positions obtained for the system at
the absolute potential minimum. The probability-density $\rho(r)$ was
normalized so that:
\begin{equation}
       \int\limits_0^\infty
                    \,d r  \; 2 \pi r \rho (r)  =   1   .
\label{6}
\end{equation}
In Fig.\,\ref{fullsim} the density $\rho(r)$ for the
oxygen nucleus is shown by a full line.
The maximum probability-density
is found at the BC site, i.e., it does not correspond
to the absolute minimum of the
potential energy at the off-center site M.
This peculiar behavior was already observed in
a previous investigation assuming an effective one-particle
potential for the impurity.\cite{ego}
The relevance of our calculation is that this
nontrivial quantum delocalization of oxygen is confirmed by a quantum
approach for the full many-body potential.
One expects that an impurity with larger mass, interacting through the
same potential as that employed for oxygen,
will approach a "classical" behavior, in the sense that those spatial
regions with larger probability-density will correspond to
regions of lower potential energy. In Fig.\,\ref{fullsim}, the function
$\rho(r)$ obtained by setting in the PI simulation
the impurity mass $m_{\rm O} = 60$ amu, is
displayed by a broken line. The maximum probability-density is found in
this case at an off-center position.
We have performed a series of simulations 
at 10 K varying the impurity mass 
in the range $m_{\rm O}$ = 16--150 amu. In 
Fig. \ref{fullsim}(b), the position of the maximum of the $\rho(r)$
curves is presented as a function of the impurity mass $m_{\rm O}$.
For a mass larger than about 50 amu we
find a "classical" regime, where the
position of the maximum of $\rho(r)$ is off-center.

We note that the assignment of a definite geometry to the $O_i$ defect
is difficult and may be meaningless, because the probability-density
function is very
broad in the plane perpendicular to the Si--Si axis. This large spatial
uncertainty is due to the zero-point motion of the impurity
and therefore it cannot be reduced by
decreasing the temperature. From the point of view of a structure with
minimum potential energy,
which is the one usually adopted in total-energy
investigations using electronic structure methods, the defect geometry
would correspond to a non-linear arrangement of the oxygen and the
nearest-neighbor Si atoms. This is of no physical significance, however, 
since the maximum probability-density (and the symmetry) is for
a linear disposition of the atoms.
It is the highly anharmonic situation found for oxygen in silicon
that leads to contradictory pictures concerning the defect geometry. This
point could be relevant also for some crystalline phases, like
$\beta$--cristobalite, where different structural models have been
proposed, and some controversy has arisen concerning the existence of
linear Si--O--Si units in this structure.\cite{liu93,swainson93}

\subsection{Decoupling into a one-particle problem}

   The results derived from the full quantum mechanical treatment of the
$O_i$ defect can be used to test the quality of several alternatives
(FLP versus AP) for the decoupling of the many-nuclei degrees-of-freedom
into a one-body potential.
In Fig. \ref{compare}(a), the "exact" probability-density $\rho(r)$
for the oxygen nucleus obtained by the full PI
simulation at 10 K is represented by open squares and
is compared to the curves derived by
four different decoupling schemes. We have analyzed three different
fixed-lattice potentials FLP and the adiabatic potential AP.
For each one of these potentials we have performed a one-particle 
PI simulation to obtain $\rho(r)$ at 10 K.
The broken line in Fig. \ref{compare}(a) is the result 
derived from the potential FLP-2,  
defined with the Si
atoms fixed at the relaxed positions obtained with the O impurity
located at the BC site.
 The maximum probability-density is found at the BC site in
agreement with the result obtained for the full potential. However, the
probability-density away from the BC site decays too
fast and the curvature of the $\rho(r)$ curve 
does not compare well with the result of the full PI
simulation. The dashed-dotted line corresponds to the potential FLP-1,
 defined with the Si atoms
fixed at the relaxed positions obtained with
the O impurity located at the off-center site
M. This potential FLP-1
leads to an even more unrealistic probability-density, as the maximum
density is found close to the off-center position M. 
The dotted line in Fig. \ref{compare}(a) was derived
from the third choice of a FLP, where the
Si atoms were fixed at their equilibrium positions
obtained from the MC trajectory generated by the full
quantum simulation of the O impurity and the Si atoms.
This potential FLP-3 gives a better agreement with the full simulation.
However, in spite of its much larger computational requirements
(it needs the results of the full quantum treatment), the improvement
is still unsatisfactory. This and the other FLP approaches discussed 
above are not able to reproduce closely
the density distribution of the impurity. At last, we present in Fig.
\ref{compare}(a) the results found for the oxygen impurity moving in the
adiabatic potential defined in the plane perpendicular to the Si--Si axis
[dotted line in  Fig. \ref{potential}(b)]. 
This $\rho(r)$ curve is shown by a full line and
agrees closely with that derived by the full quantum simulation.
We conclude that the best decoupling scheme for the $O_i$ defect,
in the plane perpendicular to the Si--Si axis,
corresponds to the AP approximation.

It is interesting to test whether the AP potential is 
also able to reproduce the
results of the full simulation shown in Fig. \ref{fullsim}(b), by a
series of one-body simulations for different impurity masses.
The results obtained in these PI MC simulations for
the position of the maximum of $\rho(r)$ are given in
Fig. \ref{compare}(b). The agreement with the data
of the full quantum simulations of Fig. \ref{fullsim}(b)
is good.  In particular, the crossover between impurity-mass ranges
with maximum probability-density at BC or off-center, 
lies at about 50 amu,
as found in the full quantum problem.
As the mass of the impurity increases, its motion with respect
to the host atoms will become relatively slower, therefore
the adiabatic potential remains a good
approximation to the many-body problem.

Finally, we compare in Tab. \ref{table1} 
some spectroscopic information,
obtained by solving numerically the two-dimensional
Schr\"odinger equation of an oxygen nucleus moving in the
AP and FLP-1 potentials, with available 
experimental data. \cite{bosomworth}
The frequencies derived from the AP potential show a closer agreement
to the experimental results than the frequencies
calculated with the FLP-1 potential.
The AP transitions are about 
40 \% larger than those derived from experiment. However, the renormalization
of the transitions to the value of the first excitation reduces 
this discrepancy to 12 \%. This deviation from experiment can be due
either to limitations in the adiabatic approximation or 
to shortcomings in the employed potential model. The main
conclusion from the data in Tab. \ref{table1}
is that the adiabatic potential appears again as a
better decoupling approach to the full many-nuclei problem
than the fixed lattice potential.

\section{CONCLUSIONS}

The definition of a defect geometry for an isolated oxygen impurity
in silicon is conditioned by a substantial deviation
from harmonicity, as the configuration with
minimum potential energy does not correspond to a maximum
in the probability-density function of the oxygen nucleus.                
The minimum energy configuration is found for
a bent Si--O--Si configuration. However, the probability-density 
function of the impurity, as derived by PI MC simulations, displays
the symmetry of a linear arrangement of atoms.
This is a case in 
which the structure associated to the minimum energy 
is not the best choice to derive some physical properties of
the defect complex. 
The anharmonicity of
the zero-point motion of the impurity
and the lattice atoms reduces the importance of the
minimum-energy structure for defining the properties of the impurity
center. 

An important property of the $O_i$ center is that the 
dynamics in the plane perpendicular to the Si--Si axis is
characterized by frequencies of about 30 cm$^{-1}$. These
extremely low frequencies, even for a relatively light impurity as
oxygen, cause that the best decoupling scheme for the 
reduction of the full many-body problem into a one-body 
treatment is given by the adiabatic potential, where for each
impurity position the Si atoms (moving faster) relax to their
equilibrium sites.

Low-energy excitations are normally associated
to localized
defect states. Some of the physics, however, is very much related to 
other dynamical phenomena like diffusion. Specifically, the problem of the
delocalization of light impurity atoms among symmetry-equivalent sites
around another atom\cite{complexes} is
essentially the same as what is found for
the diffusion of those light atoms, where the delocalization is among sites
equivalent by translational symmetry. The discussion in
this paper about the definition of a static defect
structure is partially transferable to the dynamical problem
of defining diffusion paths for quantum particles.

\acknowledgments

We thank Pedro L. de Andr\'es for critical comments on the
manuscript.
This work was supported by DGICYT (Spain) under contracts
PB93-1254 and PB92-0169.
%
% REFERENCES
%

%
\noindent
\begin{figure}[t]
\caption{Schematic representation of the atom disposition 
in the $O_i$ center. Alternative
sites for the oxygen nucleus are indicated by BC (bond-center)
and M (off-center). The absolute minimum of the employed potential model
corresponds to the off-center site M, 0.29 \AA\ away from the BC site.}
\label{sketch}
\end{figure}
\noindent
\begin{figure}[t]
\caption{(a) Potential energy for the nuclear motion shown versus the O distance
to the BC site within the plane perpendicular to the Si--Si axis,
and versus the Si--Si distance between host atoms nearest-neighbors of O.
The contour interval is 3 meV.
 (b) Three different one-dimensional cuts are presented:
(i) for the Si atoms fixed at the positions of the
absolute potential-minimum (fixed-lattice potential FLP-1, broken line); 
(ii) for the Si atoms fixed at their relaxed positions for O at the BC
site (FLP-2, full line) , and 
(iii) for the Si atoms relaxing at each O position (adiabatic
potential, AP,  dotted line).
 (i) and (ii) are cuts of the potential surface in (a), in vertical planes
parallel to the $d$(O--BC) axis. }
\label{potential}
\end{figure}
\noindent
\begin{figure}[t]
\caption{(a) Probability-density $\rho(r)$ for the O nucleus
obtained in the full PI simulation at 10 K. 
$r$ is the distance from the impurity
to the Si--Si axis. The full line corresponds to $m_{\rm O}$ = 16 amu
and the broken line to $m_{\rm O}$ = 60 amu.
(b) Distance from the probability-density maximum to the Si--Si axis,
 as a function of the impurity mass. The broken line is a guide to the
eye.}
\label{fullsim}
\end{figure}
\noindent
\begin{figure}[t]
\caption{(a) Comparison of probability-density curves
of the O nucleus for different one-particle
potentials with that derived by the many-particle simulation
(open squares). The broken, dotted, and dashed-dotted lines
correspond to different FLP models (see text), while the full line
is the result of the adiabatic potential (AP).
(b) Position of the maximum in the probability-density
 $\rho(r)$ corresponding to the adiabatic potential,
as a function of the impurity mass. The broken line is a guide to the
eye.}
\label{compare}
\end{figure}
\newpage
\narrowtext
\begin{table}
\caption{
Observed and calculated FIR transitions (cm$^{-1}$) of
oxygen in silicon.
The calculated values correspond to the AP and
FLP-1 potentials.
In parenthesis are given the relative values with respect to the first
transition. The observed values are taken from Ref. [1]. }
\begin{tabular}{lccc}
Transition  & obs.  & AP  & FLP-1 \\
\tableline
$|0,0 \rangle \rightarrow     |1,\pm 1\rangle$ & 29.3 \, (1)  & 
37.1  \, (1) & 24.1 \, (1)  \\
$|1,\pm 1 \rangle \rightarrow |2,\pm 2\rangle$ & 37.8 \, (1.3) & 
51.6 \, (1.4) & 51.6 \, (2.1) \\
$|2,\pm 2 \rangle \rightarrow |3,\pm 3\rangle$ & 43.3 \, (1.5) & 
62.9 \, (1.7) &  70.9 \, (2.9) \\
$|1,\pm 1 \rangle \rightarrow |2,    0\rangle$ & 49.0 \, (1.7) &
72.6 \, (2.0) &  140.3 \, (5.8) \\
\end{tabular}
\label{table1}
\end{table}

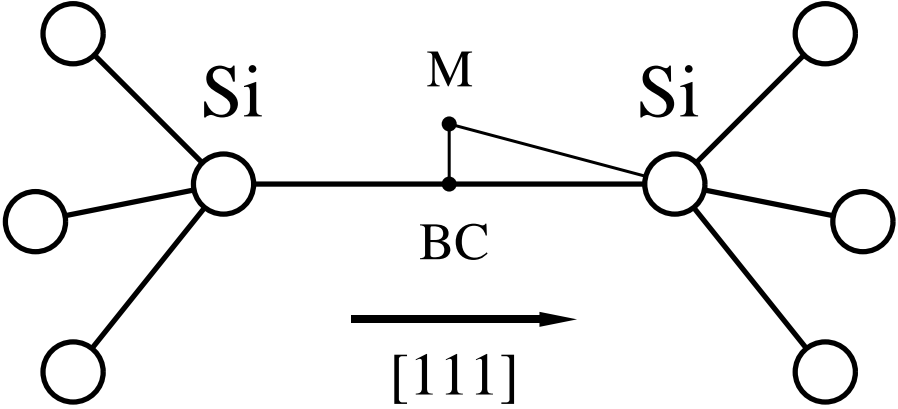
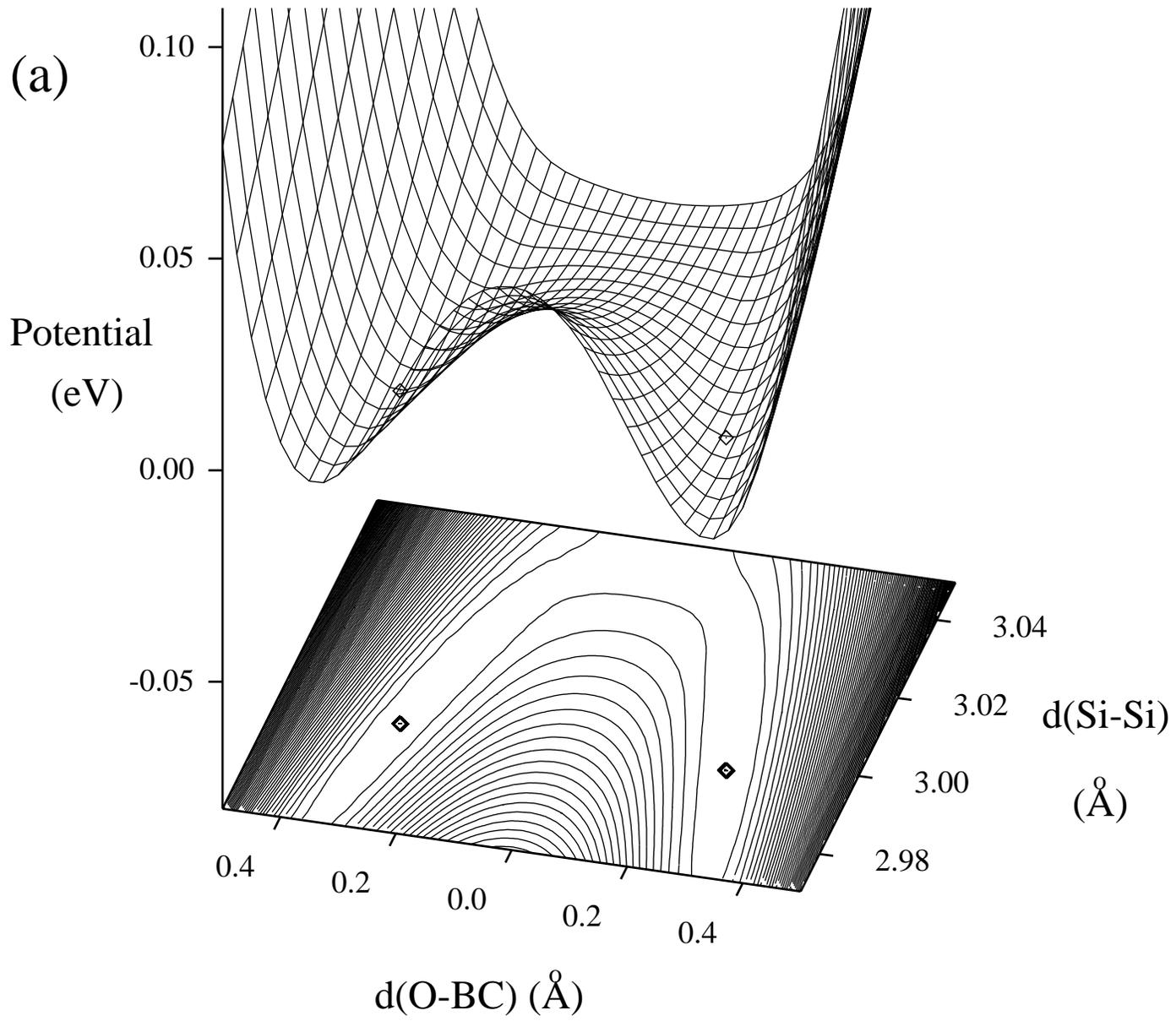
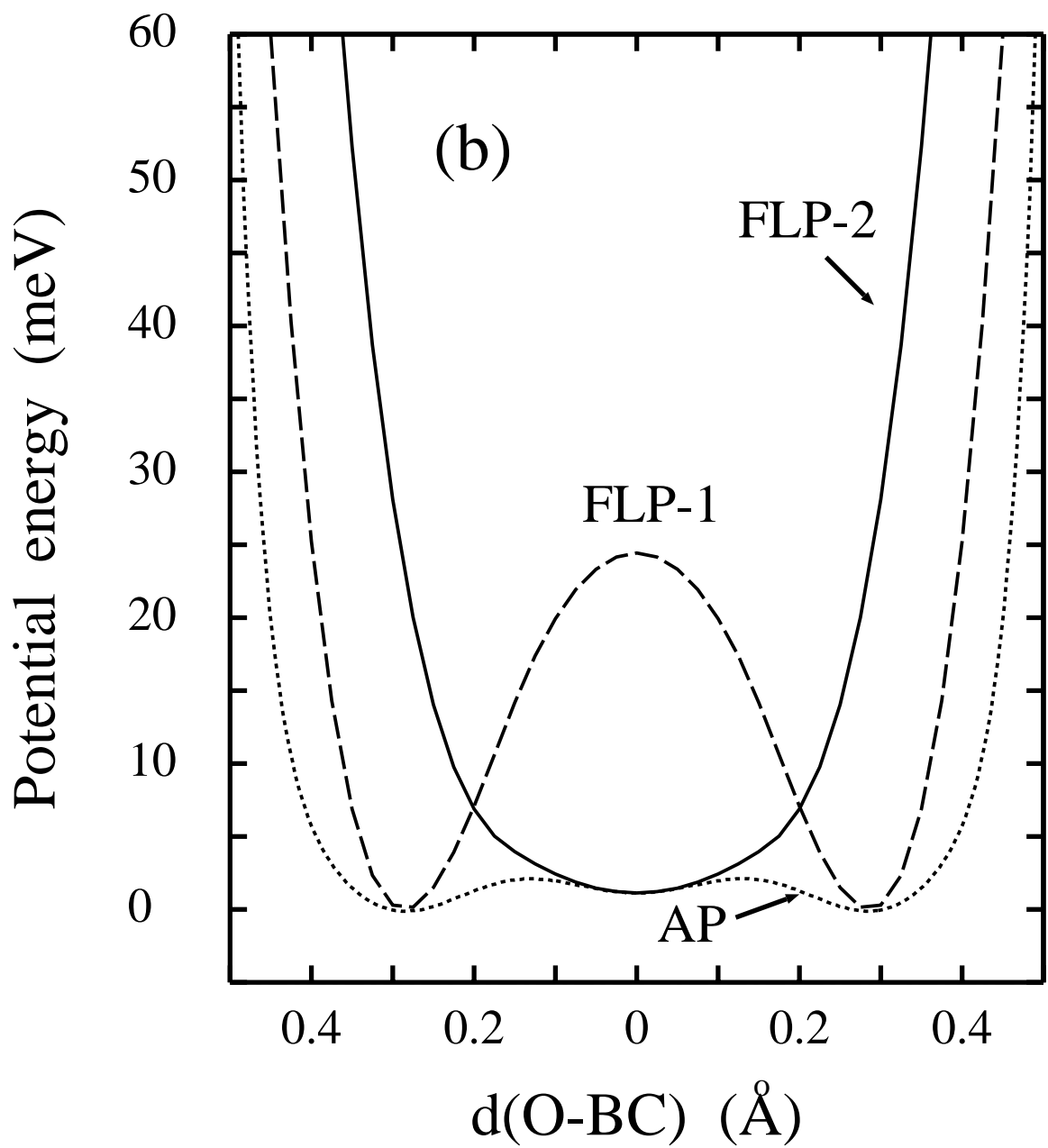
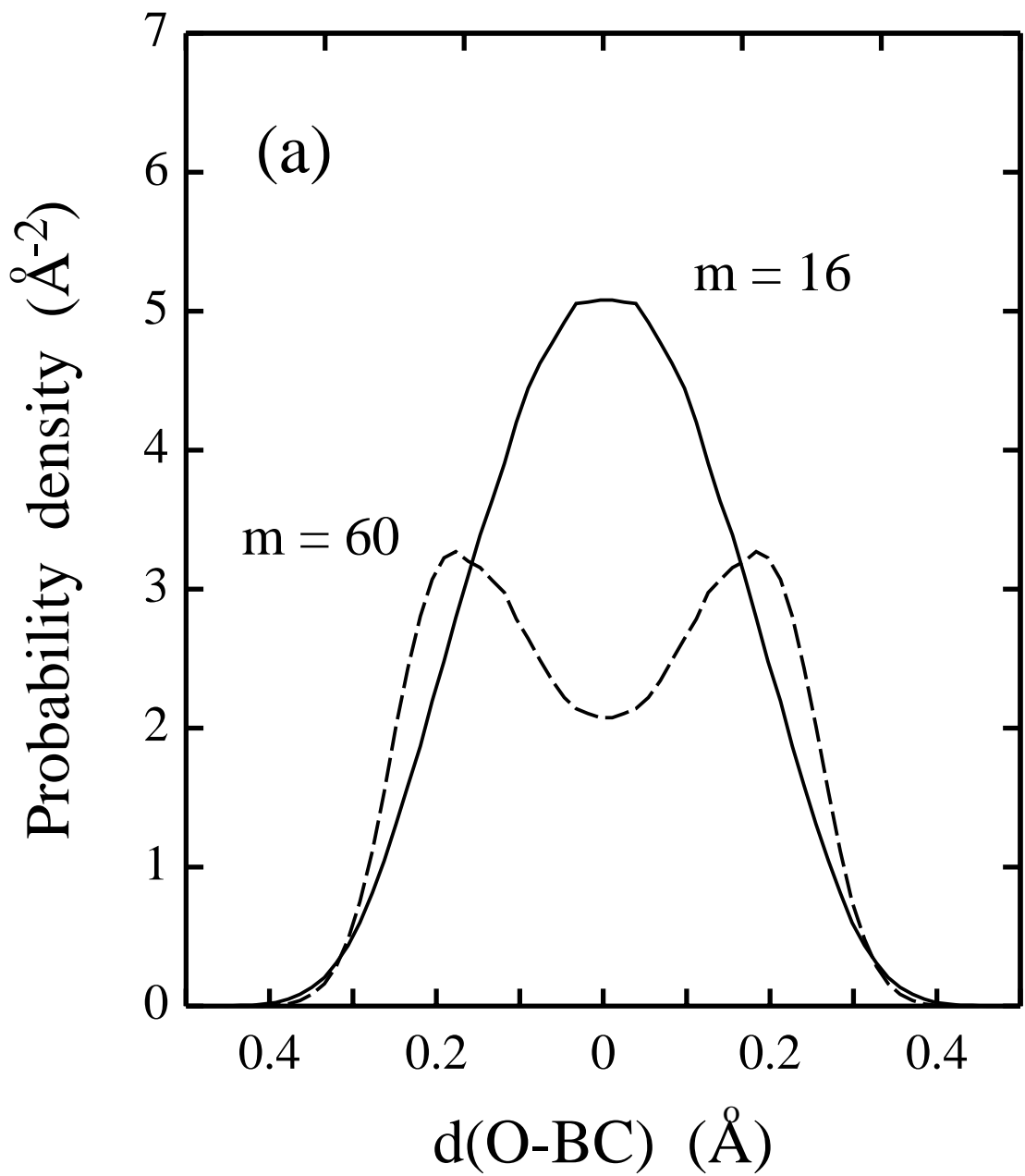
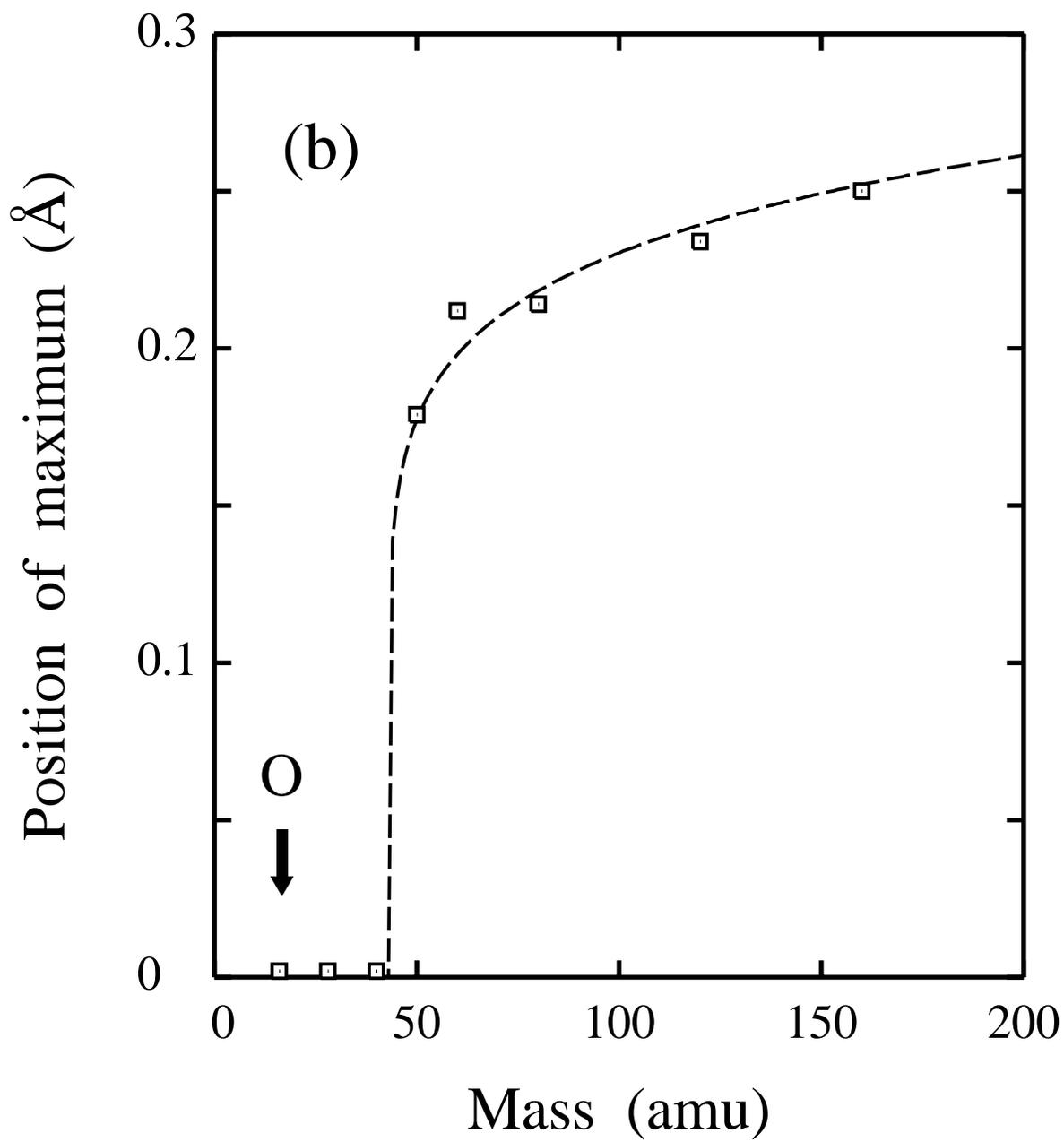
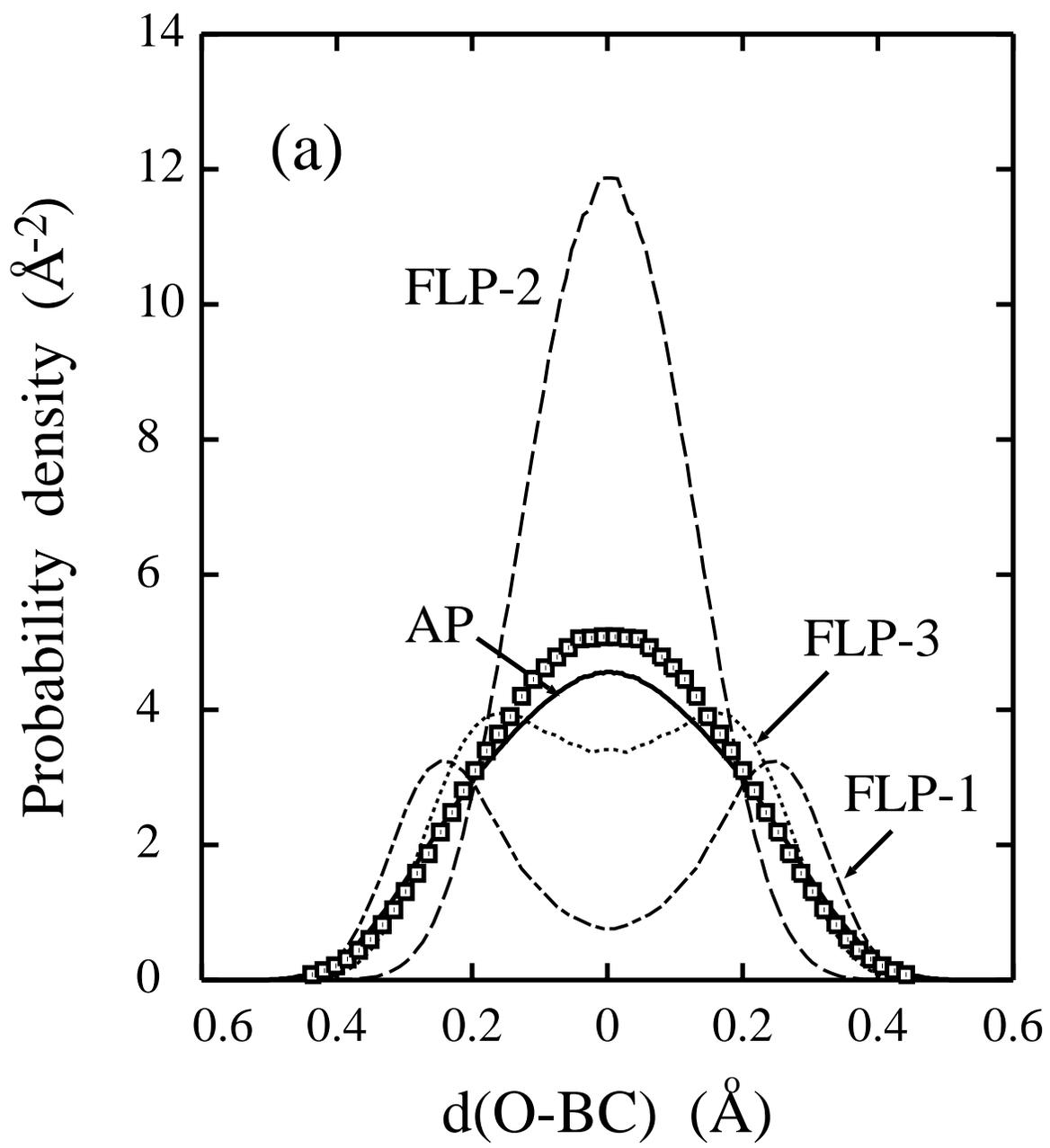
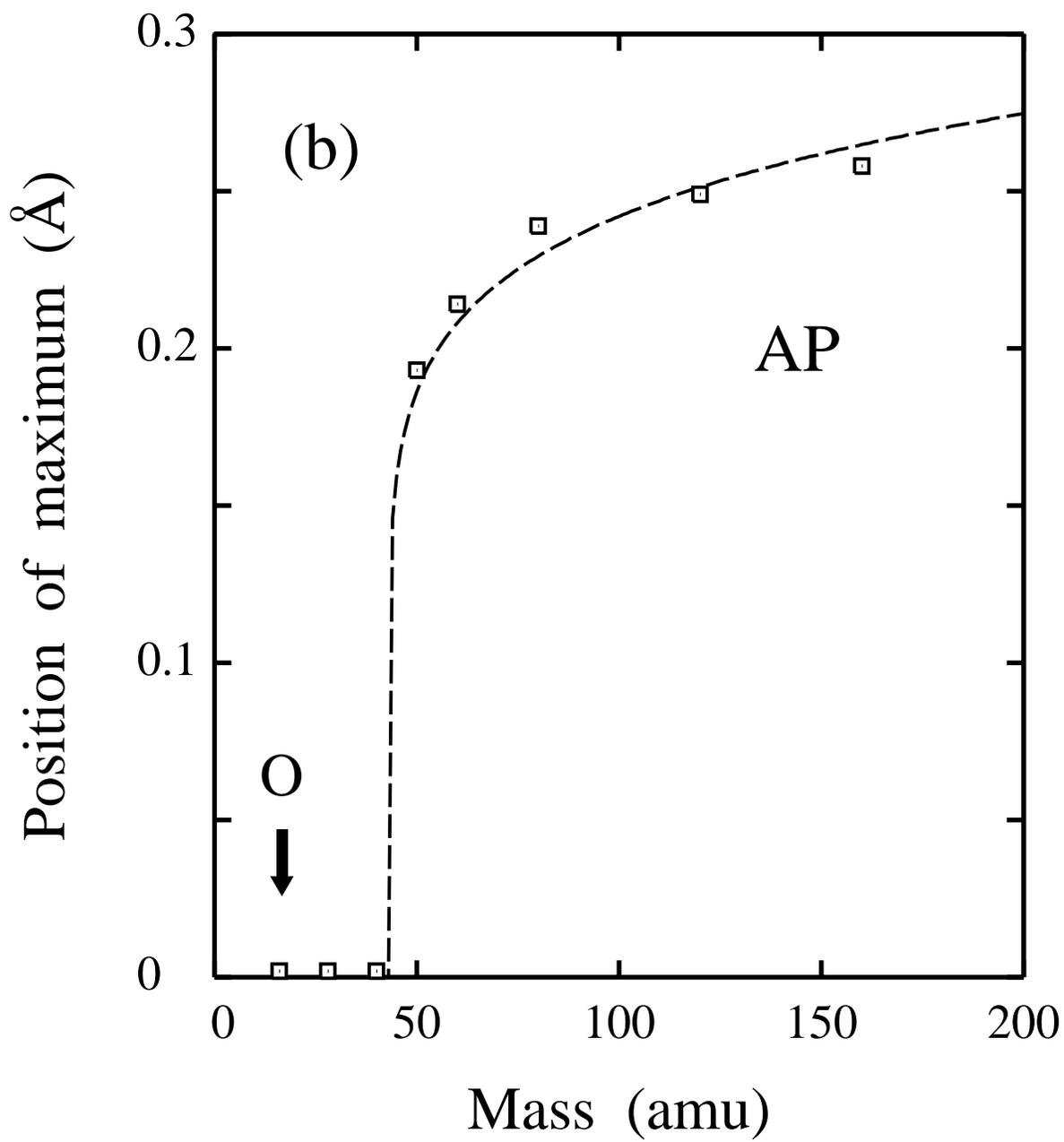
\end{document}